\newcommand\deq{\delta Q_{rev}}
\newcommand\pa{\partial}

\newcommand\beq{\begin{equation}}
\newcommand\eeq{\end{equation}}
\newcommand\beqnl{\begin{eqnarray}}
\newcommand\beqna{\begin{eqnarray*}}
\newcommand\eeqna{\end{eqnarray*}}
\newcommand\eeqnl{\end{eqnarray}}

 \def\NN{\hbox{\sf I\kern-.13em\hbox{N}}}
 \def\HH{\hbox{\sf I\kern-.13em\hbox{H}}}
 \def\DD{\hbox{\sf I\kern-.13em\hbox{D}}}
 \def\RR{\hbox{\sf I\kern-.14em\hbox{R}}}
 \def\CC{\hbox{\sf I\kern-.44em\hbox{C}}}
 \def\ZZ{{\hbox{\sf Z\kern-.43emZ}}}
 \def\QQ{\hbox{\sf C\kern -.48emQ}}
 \def\Cc{\hbox{\sf C\kern -.47em {\raise .48ex \hbox{$\scriptscriptstyle |$}}
   \kern-.5em {\raise .48ex \hbox{$\scriptscriptstyle |$}} }}
 \def\Qq{\hbox{\sf Q\kern -.57em {\raise .48ex \hbox{$\scriptscriptstyle |$}}
   \kern-.55em {\raise .48ex \hbox{$\scriptscriptstyle |$}} }}

\documentstyle[prl,aps,multicol,psfig]{revtex}
\begin{document}
\draft
%@twocolumnfalse\endcsname
%-------------------------------------------------------------------------
%\newcommand\bfg[1]{\begin{figure}\vspace{#1cm}}
%\newcommand\efg{\end{figure}}
%-------------------------------------------------------------------------

%\rightline{Draft}
%\rightline{gr-qc/0210031}
\vskip1pc

\title{Notes on Quasi-Homogeneous Functions in Thermodynamics}
\author{F. Belgiorno\footnote{E-mail address: belgiorno@mi.infn.it}}
\address{Dipartimento di Fisica, Universit\`a degli Studi di Milano, 
Via Celoria 16, 20133 Milano, Italy}%, and\\
%I.N.F.N., sezione di Milano, Italy}
\date{\today}
\maketitle

\begin{abstract}

A special kind of quasi-homogeneity occurring in 
thermodynamic potentials of standard thermodynamics 
is pointed out. Some formal consequences are also 
discussed.

\end{abstract}

%, 04.70.Dy}
%\begin{multicols}{2}
%\narrowtext

\section{introduction}
\label{qotde}

Quasi-homogeneous functions have been introduced in the framework 
of standard thermodynamics with the aim to studying scaling and 
universality near the critical point \cite{hankey,neff}. A common 
synonymous of ``quasi-homogeneous function'' is 
``generalized homogeneous function'' [see e.g. \cite{hankey,neff,aczel}].  
We wish to point out here that quasi-homogeneity can be an useful tool 
in the framework of standard thermodynamics, when one considers 
intensive variables as independent variables for the equilibrium 
thermodynamics description of a system. In fact, homogeneity for 
the fundamental equation in the entropy representation [and in the 
energy representation] is well-defined in terms of the standard 
Euler theorem for homogeneous functions \cite{callen}. 
One simply defines the 
standard Euler operator (sometimes called also Liouville operator) 
and requires the entropy [energy] to be an homogeneous function 
of degree one. When the other thermodynamic potentials which 
are obtained from the entropy [energy] are taken into account 
by means of suitable Legendre transformations, then part of the 
independent variables are intensive \cite{callen}. 
The thermodynamic potentials 
are still homogeneous of degree one in the extensive independent 
variables, but a different rescaling is appropriate for the 
independent variables. For example, let us consider the Gibbs 
potential $G(T,p,N)$ for a system which is described by means 
of three independent variables $T,p,N$. $G$ is homogeneous 
of degree one when the system is rescaled by $\lambda$, such 
a rescaling corresponding only to a rescaling $N\to \lambda N$, 
because $T$ and $p$ are intensive and remain unchanged under 
rescaling of the system.  
This is evident because, as it is well known, one has 
$G=\mu (p,T) N$, where $\mu$ is the chemical potential. 
Actually, one could also define $G$ as a quasi-homogeneous 
function of degree one with weights $(0,0,1)$. Then the behavior 
under scaling is better defined. 
A mathematical treatment of the same problem is found in Ref. 
\cite{honig}. The approach we present here is characterized by the 
more general setting allowed by the technology of quasi-homogeneous 
functions; the sections on the Gibbs-Duhem equation and the 
on Pfaffian forms contain a further analysis of some formal aspects of 
standard thermodynamics.
 
\section{quasi-homogeneous functions and thermodynamics}

Given a set of real 
coordinates $x^1,\ldots,x^n$ and a set of weights 
$\mbox{\boldmath $\alpha$}\equiv (\alpha_1,\ldots,\alpha_n) \in \RR^n$, 
a function $F(x^1,\ldots,x^n)$ is quasi-homogeneous of degree $r$ 
and type $\mbox{\boldmath $\alpha$}$ \cite{anosov} if, under dilatations by 
a scale factor $\lambda>0$ one finds
\beq
F(\lambda^{\alpha_1}\; x^1,\ldots,\lambda^{\alpha_1}\;
x^n) = \lambda^r\; F(x^1,\ldots,x^n).
\eeq
A differentiable quasi-homogeneous function satisfies 
a generalized Euler identity: 
\beq
D\; F=r\; F,
\label{geul}
\eeq 
where $D$ is 
the Euler vector field  
\beq
D\equiv \alpha_1\; x^1\; \frac{\pa}{\pa x^1}+\ldots+
\alpha_n\; x^n\; \frac{\pa}{\pa x^n}.
\label{veuler}
\eeq
Notice that (\ref{geul}) is a necessary and sufficient condition for a 
differentiable function to be quasi-homogeneous \cite{anosov}. 
It is also interesting to define quasi-homogeneous 
Pfaffain forms. A Pfaffian form 
\beq
\omega=\sum_{i=1}^n\; \omega_i(x)\; dx^i
\eeq
is quasi-homogeneous of degree $r\in \RR$ if, 
under the scaling 
\beq
x^1,\ldots,x^n \to 
\lambda^{\alpha_1}\; x^1,\ldots,\lambda^{\alpha_n}\; x^n
\eeq
one finds
\beq
\omega \to \lambda^r\; \omega.
\eeq
This happens if and only if the degree of quasi-homogeneity 
deg$(\omega_i (x))$ of $\omega_i(x)$ is such that 
deg$(\omega_i (x))=r-\alpha_i\quad \forall i=1,\ldots,n$. 
For a discussion about quasi-homogeneity and for further references, 
see \cite{qobh}.

\subsection{quasi-homogeneous potentials in standard thermodynamics}

Let us consider a thermodynamic potential 
$R(y^1,\ldots,y^k,x^{k+1},\ldots,x^n)$ depending on 
$k$ intensive variables $y^1,\ldots,y^k$ and $n-k$ extensive 
variables $x^{k+1},\ldots,x^n$. $R$ is required to be quasi-homogeneous 
of degree 1 and its type is 
\beq
\mbox{\boldmath $\alpha$}=(\underbrace{0,\ldots,0}_{k},
\underbrace{1,\ldots,1}_{n-k}).
\eeq
Then, one has 
\beq
R=\sum_{i=k+1}^{n} x^i\; 
\frac{\pa R}{\pa x^i}.
\label{eulext}
\eeq
This expression of the thermodynamic potentials is well-known, 
it is sometimes referred to as the identity satisfied by the 
potentials at fixed intensive variables \cite{beattie}. 
A treatment on a mathematical ground of the same topic is found in 
Ref. \cite{honig}.  
It is evident that, in order to ensure that $R$ is a degree one 
quasi-homogeneous function, the intensive variables can be at most 
$n-1$, in which case (cf. also the following section)
\beq
R=x^n\; 
\frac{\pa R}{\pa x^n}\equiv x^n\; r(y^1,\ldots,y^{n-1}),
\label{onext}
\eeq
where $r(y^1,\ldots,y^{n-1})$ is of degree zero.\\ 
We recall that, given the fundamental equation of thermodynamics in 
the energy [entropy] representation, one can obtain other fundamental 
equations by means of the Legendre transform \cite{callen}. It is 
easy to show that:\\ 
\\
{\sf the Legendre transform with respect to a variable 
of weight $\alpha$ of a quasi-homogeneous function of degree $r$ 
is a quasi-homogeneous function of degree $r$ with the weight 
$\alpha$ changed into the weight $r-\alpha$ of the Legendre-conjugate 
variable} (theorem 2 of \cite{hankey}).\\ 
\\
Moreover,\\ 
\\
{\sf the partial derivative with respect 
to  a variable 
of weight $\alpha$ of a quasi-homogeneous function $R$ of degree $r$ is 
a quasi-homogeneous function of degree $r-\alpha$ having the same type  
as $R$} 
(theorem 1 of \cite{hankey}). See also \cite{qobh}.\\ 
\\
These results allow to justify 
easily the following examples.\\ 
For the free energy $F(T,V,N)$, one has $F=U-TS$, thus $F$ is  
a quasi-homogeneous function of degree 1 and of weights 
$(0,1,1)$, and 
\beq
F(T,V,N)=V\; \frac{\pa F}{\pa V}+N\; \frac{\pa F}{\pa N}.
\eeq
Analogously, 
\beq
S(T,V,N)=V\; \frac{\pa S}{\pa V}+N\; \frac{\pa S}{\pa N}.
\eeq
[In fact, $S=-\pa F/\pa T$ and theorem 1 of \cite{hankey} can 
be applied]. 
Moreover, given $S(T,p,N)$, one has
\beq
S(T,p,N)=N\; \frac{\pa S}{\pa N}.
\eeq
In concluding this section, we point out that the distinction between 
degree and weights of thermodynamic variables is somehow artificial, 
a degree becoming a weight if the thermodynamic variable is changed into 
an independent variable (e.g., the degree zero of the pressure becomes 
a weight zero when $p$ is an independent variable).

\section{Gibbs-Duhem equations}

Herein we take into account the Gibbs-Duhem equations. 
Cf. also \cite{honig}.
Let us define 
\beq
R_i\equiv \frac{\pa R}{\pa x^i};\qquad R_a\equiv \frac{\pa R}{\pa y^a}
\eeq
one has 
\beq
dR=\sum_{a=1}^{k}\; R_a\; dy^a+ \sum_{i=k+1}^{n}\; R_i\; dx^i.
\eeq
On the other hand, one obtains from (\ref{eulext})
\beq
dR=\sum_{i=k+1}^{n}\; R_i\; dx^i+ \sum_{i=k+1}^{n}\; x^i\; dR_i.
\eeq
The GD equation is then 
\beq
\sum_{a=1}^{k}\; R_a\; dy^a-\sum_{i=k+1}^{n}\; x^i\; dR_i=0.
\label{gdeq}
\eeq
This equation is related with the quasy-homogeneity symmetry 
of the potential. Let us define the Euler operator 
\beq
X\equiv  \sum_{i=k+1}^{n}\; x^i\; \frac{\pa}{\pa x^i}.
\eeq
Let us also define a 1-form 
\beq
\omega_R\equiv \sum_{a=1}^{k}\; R_a\; dy^a+ \sum_{i=k+1}^{n}\; R_i\; dx^i
\eeq
where $R_a$ are quasi-homogeneous functions of degree one $X\; R_a=R_a$ and 
the $R_i$ are quasi-homogeneous functions of degree zero $X\; R_i=0$. 
Then $\omega_R$ is a quasi-homogeneous 1-form of degree one, in the 
sense that it satisfies $L_X\; \omega_R=\omega_R$, where $L_X$ is 
the Lie derivative associated with $X$. One can also define a 
function 
\beq
R\equiv i_X \omega_R,
\eeq
where $i_X$ is the standard contraction operator. As a consequence, 
one finds 
\beq
dR = d (i_X \omega_R)=-i_X\; d\omega_R+ L_X\; \omega_R=-i_X\; d\omega_R+
\omega_R
\label{eqgd}
\eeq
If $\omega_R$ is a closed 1-form (and then, exact in the convex 
thermodynamic domain), then $d\omega_R=0$ and $dR=\omega_R$, i.e. 
$R$ is the potential associated with $\omega_R$. Notice also that 
in the latter case one finds  
\beq
-i_X\; d\omega_R=0
\label{gdgeo}
\eeq
which corresponds to the Gibbs-Duhem equation. In fact, one has 
\beq
d\omega_R=
\sum_{a=1}^{k}\; dR_a\wedge dy^a+ \sum_{i=k+1}^{n}\; dR_i\wedge dx^i
\eeq
and 
\beqnl
i_X\; d\omega_R &=& \sum_{a=1}^{k}\; (i_X\; dR_a)\;  dy^a-
\sum_{a=1}^{k}\; dR_a\; (i_X\; dy^a)\cr
&+&\sum_{i=k+1}^{n}\; (i_X\; dR_i)\;  dx^i-
\sum_{i=k+1}^{n}\; dR_i\; (i_X\; dx^i)\cr
&=&\sum_{a=1}^{k}\; R_a\;  dy^a-
\sum_{i=k+1}^{n}\; x^i\; dR_i=0,
\eeqnl
where $i_X\; dR_a=X\; R_a=R_a$, and $i_X\; dR_i=X\; R_i=0$.\\ 
The converse is also true, i.e., if (\ref{gdgeo}) is satisfied 
then from (\ref{eqgd}) follows that $\omega_R$ is closed.\\
The GD equation is then satisfied because of the equality of the 
mixed second derivatives of $R$ (Schwartz theorem) and 
because of the quasi-homogeneous symmetry. 
In fact, by defining $Q_{\alpha\beta}$ the matrix of the second 
partial derivatives of $R$, one finds
\beq
\sum_{i=k+1}^{n}\; x^i\; dR_i = 
\sum_{a=1}^{k}\; \sum_{i=k+1}^{n}\; x^i\; Q_{ia}\; dy^a+
\sum_{j=k+1}^{n}\; \sum_{i=k+1}^{n}\; x^i\; Q_{ij}\; dx^j.
\eeq
Then the Gibbs-Duhem equation (\ref{gdeq}) is equivalent to 
\beqnl
\sum_{a=1}^{k}\; \sum_{i=k+1}^{n}\; x^i\; Q_{ia}\; dy^a &=& 
\sum_{a=1}^{k}\; R_a\; dy^a \label{gdp}\\
\sum_{j=k+1}^{n}\; \sum_{i=k+1}^{n}\; x^i\; Q_{ij}\; dx^j &=& 0. \label{gds}
\eeqnl
The former formula (\ref{gdp}) is implemented if both Schwartz theorem and 
the quasi-homogeneous symmetry are implemented. In fact, 
\beqnl
\sum_{a=1}^{k}\; \sum_{i=k+1}^{n}\; x^i\; Q_{ia}\; dy^a &=& 
\sum_{a=1}^{k}\; \sum_{i=k+1}^{n}\; x^i\; \left( \frac{\pa}{\pa x^i}
\; \frac{\pa}{\pa y^a}\; R\right)\; dy^a\cr 
&=&\sum_{a=1}^{k}\; \frac{\pa}{\pa y^a}\; 
\left(  \sum_{i=k+1}^{n}\; x^i\; \frac{\pa}{\pa x^i}\;  R\right)\; dy^a\cr
&=&\sum_{a=1}^{k}\; \frac{\pa R}{\pa y^a}\; dy^a=\sum_{a=1}^{k}\; R_a\; dy^a.
\eeqnl
Also (\ref{gds})  is implemented, in fact
\beqnl
\sum_{j=k+1}^{n}\; \sum_{i=k+1}^{n}\; x^i\; Q_{ij}\; dx^j &=& 
\sum_{j=k+1}^{n}\; \sum_{i=k+1}^{n}\; x^i\; \left( \frac{\pa}{\pa x^i}
\; \frac{\pa}{\pa x^j}\; R\right)\; dx^j\\
&=& 
\sum_{j=k+1}^{n}\; \left(  
\sum_{i=k+1}^{n}\; x^i\; \frac{\pa}{\pa x^i}\; \frac{\pa R}{\pa x^j} 
\right)\; dx^j,
\eeqnl
and the latter is zero because $\pa R/\pa x^i$ are functions of degree 
zero for all $i=k+1,\ldots,n$. 

\section{Pfaffian form $\deq$}

Let us consider the Pfaffian form $\deq$ for a system described by 
$(T,V,N)$, where $T$ is the absolute temperature; one has 
\beq
\deq = C_{VN}(T) dT+a(T,V,N) dV + b(T,V,N) dN.
\eeq
$\deq$ has to be integrable, i.e., it satisfies $\deq \wedge d(\deq)=0$, 
and it is known that 
$T$ is an integrating factor for $\deq$, with 
\beq
\frac{\deq}{T}=dS.
\label{des}
\eeq
Then, one finds that 
\beq
\frac{\deq}{T} = \frac{C_{VN}(T)}{T} dT+ \frac{a(T,V,N)}{T} dV + 
\frac{b(T,V,N)}{T} dN
\eeq
is exact and a potential is given by 
\beq
S=\frac{a(T,V,N)}{T}\; V + 
\frac{b(T,V,N)}{T}\; N.
\label{qos}
\eeq
Notice that the quasi-homogeneity of degree one of $S$ is 
the tool allowing to obtain this result. It is ``trivial'' that 
$S$ is the potential associated with $\deq/T$, it is less 
trivial that its ``homogeneity'' leads to (\ref{qos}).  
For a proof, see the appendix. 
$\deq$ is quasi-homogeneous of degree one and weights $(0,1,1)$. 
From the theory of quasi-homogeneous integrable Pfaffian forms 
\cite{qobh}, it is known that an integrating factor is also given by 
\beq
f = a(T,V,N)\; V+ b(T,V,N)\; N.
\eeq
The proof is found in Ref. \cite{qobh}. It is evident that 
\beq
f= T S.
\eeq
Analogously, one can 
consider $(T,p,N)$ as independent variables
\beq
\deq = C_{pN}(T) dT+\eta (T,p,N) dp + \zeta (T,p,N) dN,
\eeq
in which case
\beq
f=\zeta(T,p)\; N = T S.
\eeq

\appendix
\section{potentials of exact quasi-homogeneous Pfaffian forms}

We show that, if 
\beq
\omega = \sum_{a=1}^{k}\; B_a (y^1,\ldots,y^k,x^{k+1},\ldots,x^n)\; 
dy^a+\sum_{i=k+1}^{n}\; B_i (y^1,\ldots,y^k,x^{k+1},\ldots,x^n)\; 
dx^i
\eeq
is a $C^2$ exact quasi-homogeneous Pfaffian form of degree one, with 
$B_a,x^i$ quasi-homogeneous of degree one and 
$B_i,y^a$ quasi-homogeneous of degree zero with respect to 
the Euler operator
\beq
Y=\sum_{i=k+1}^n\; x^i \frac{\pa }{\pa x^i},
\eeq
then 
\beq
P(y^1,\ldots,y^k,x^{k+1},\ldots,x^n) \equiv 
\sum_{i=k+1}^{n}\; B_i (y^1,\ldots,y^k,x^{k+1},\ldots,x^n)\; x^i
\eeq
is a potential associated with $\omega$. In fact, let us consider 
\beqnl
dP &=& \sum_{i=k+1}^{n}\; B_i (y^1,\ldots,y^k,x^{k+1},\ldots,x^n)\; dx^i+
\sum_{i=k+1}^{n}\; x^i\; dB_i (y^1,\ldots,y^k,x^{k+1},\ldots,x^n)\\
&=& \sum_{i=k+1}^{n}\; B_i (y^1,\ldots,y^k,x^{k+1},\ldots,x^n)\; dx^i+
\sum_{i=k+1}^{n}\; x^i\; \sum_{j=k+1}^{n}\; 
\frac{\pa B_i}{\pa x^j} (y^1,\ldots,y^k,x^{k+1},\ldots,x^n)\; dx^j\\
&+& 
\sum_{i=k+1}^{n}\; x^i\; \sum_{a=1}^{k}\; 
\frac{\pa B_i}{\pa y^a} (y^1,\ldots,y^k,x^{k+1},\ldots,x^n)\; dy^a.
\eeqnl
The exactness of the Pfaffian form $\omega$ implies that $d\omega=0$ 
and, in particular 
\beqnl
\frac{\pa B_i}{\pa y^a}&=&\frac{\pa B_a}{\pa x^i}\qquad a=1,\ldots,k;\ 
i=k+1,\ldots,n,\\
\frac{\pa B_i}{\pa x^j}&=&\frac{\pa B_j}{\pa x^i}\qquad i,j=k+1,\ldots,n 
\eeqnl
Then, one obtains
\beqnl
\sum_{i=k+1}^{n}\; x^i\; \sum_{a=1}^{k}\; 
\frac{\pa B_i}{\pa y^a} (y^1,\ldots,y^k,x^{k+1},\ldots,x^n)\; dy^a
&=&
\sum_{i=k+1}^{n}\; x^i\; \sum_{a=1}^{k}\; 
\frac{\pa B_a}{\pa x^i} (y^1,\ldots,y^k,x^{k+1},\ldots,x^n)\; dy^a\\
&=&
\sum_{a=1}^{k}\;
\left(
\sum_{i=k+1}^{n}\; x^i\; \frac{\pa }{\pa x^i}\; 
B_a (y^1,\ldots,y^k,x^{k+1},\ldots,x^n) \right) dy^a\\
&=&\sum_{j=1}^{k}\; B_a (y^1,\ldots,y^k,x^{k+1},\ldots,x^n)\; dy^a,
\eeqnl
because each $B_a$ is quasi-homogeneous of degree one. 
On the other hand, one has 
\beqnl
\sum_{i=k+1}^{n}\; x^i\; \sum_{j=k+1}^{n}\; 
\frac{\pa B_i}{\pa x^j} (y^1,\ldots,y^k,x^{k+1},\ldots,x^n)\; dx^j
&=&
\sum_{i=k+1}^{n}\; x^i\; \sum_{j=k+1}^{n}\; 
\frac{\pa B_j}{\pa x^i} (y^1,\ldots,y^k,x^{k+1},\ldots,x^n)\; dx^j\\
&=&
\sum_{j=k+1}^{n}\; 
\left(\sum_{i=k+1}^{n}\; x^i\; \frac{\pa }{\pa x^i}\; 
B_j (y^1,\ldots,y^k,x^{k+1},\ldots,x^n) \right)\; dx^j\\
&=& 0,
\eeqnl
because each $B_i$ is quasi-homogeneous of degree zero.

\end{document}